\documentclass[12pt,onecolumn]{IEEEtran}

\usepackage{cite}
\usepackage{amsfonts,amsmath,amssymb,amsthm,graphicx}
\usepackage[]{authblk}

\newtheorem{dfn}{Definition}
\newtheorem{lm}{Lemma}

\newtheorem{crl}{Corollary}

\newtheorem{thm}{Theorem}%

\newtheorem{rmk}{Remark}

\newcommand{\scprod}[2]{\left<#1|#2\right>}
\newcommand{\ket}[1]{\left|#1\right>}
\newcommand{\bra}[1]{\left<#1\right|}
\newcommand{\tr}[1]{\text{Tr}\left(#1\right)}
\newcommand{\RR}{{\mathbb R}}

\newcommand{\NN}{{\mathbb N}}

\newcommand{\norm}[1]{\lVert#1\rVert}

\newcommand{\dv}[2]{{\frac{\partial #1}{\partial #2}}}

\newcommand{\dotex}{\frac{\partial}{\partial t}}

\newcommand{\proj}[1]{\left|#1\right>\left<#1\right|}

\title{Hamiltonian identification through enhanced observability utilizing quantum control}

\author[1]{Zaki Leghtas$^\dag$\footnote{$^\dag $Author to whom correspondance should be addressed}}
\author[2]{Gabriel Turinici}
\author[3]{Herschel Rabitz}
\author[4]{Pierre Rouchon}

\affil[1]{\small{INRIA Paris-Rocquencourt
Domaine de Voluceau, BP105
78153 Le Chesnay Cedex, France. E-mail: zaki.leghtas@inria.fr}
}
\affil[2]{\small{CEREMADE, Universit\'e Paris Dauphine, Place du Marechal De Lattre De Tassigny,
75775 Paris Cedex 16, France. E-mail: gabriel.turinici@dauphine.fr}
}
\affil[3]{\small{ Department of Chemistry, Princeton University, Frick Laboratory,
Princeton, NJ 08544, USA. E-mail: hrabitz@princeton.edu}
}
\affil[4]{\small{Mines-ParisTech, Centre Automatique et Syst\`emes,
60, boulevard Saint-Michel
75272 Paris Cedex, France. E-mail: pierre.rouchon@mines-paristech.fr}
}

\begin{document}

\bibliographystyle{IEEEtran}

\maketitle


\begin{abstract}
This paper considers Hamiltonian identification for a controllable quantum system with non-degenerate transitions and  a known initial state.  We assume to have at our disposal a single scalar control input and the population measure of only one state at an (arbitrarily large) final time T. We prove that the quantum dipole moment matrix is locally observable in the following sense:  for any two close but distinct dipole moment matrices, we construct discriminating controls giving  two different measurements.
Such discriminating controls are constructed to have three well defined temporal components, as inspired by Ramsey interferometry.
This result suggests that what may appear at first to be very restrictive measurements are actually rich for identification, when  combined with well designed discriminating controls, to uniquely identify the complete dipole moment of such systems.  The assessment supports the employment of quantum control as a promising means to achieve high quality identification of a Hamiltonian.
\end{abstract}


\section{Introduction}
\label{sec:introduction}
Quantum control has been receiving increasing attention \cite{Brif-Chakrabarti-Rabitz-NJPhys_2010} and one of its promising applications is to Hamiltonian identification \cite{Warren-Rabitz-Dahleh-Science_1993} by using the ability to actively control a quantum system as a means to gain information about the underlying Hamiltonian governing its dynamics. The underlying premise is that controls may be found which make the measurements not only robust to noise but also  highly sensitive to the unknown parameters in the Hamiltonian. Hence, although the performance of laboratory measurements may be constrained, the ability to control a quantum system has the prospect of turning this data into a rich source of information on the system's Hamiltonian.

In this paper, we consider the problem of identifying the dipole moment matrix (which is assumed to be real) of an $N-$level quantum system, initialized to a known state (ground state),  from a single population measurement at one arbitrarily large time $T$. We suppose an ability to freely control the system with a time dependent electric field $\epsilon(t)$. The measurements are obtained by (i) initializing the system's state at a known state $\ket{i}$, (ii) controlling the system with an electric field $\epsilon_k(t)$ and (iii) measuring the population of one state $\ket{f}$. This may be repeated for many controls $(\epsilon_k)_k$. A typical setting is for the system under study to be composed of a low pressure cloud of atoms or molecules with the interaction between the system and the control taking place on a time scale of $\sim 100 fs$ with the measurement of a population performed on a time scale of $\sim 100 ns$. The large difference of time scales ($1fs=10^{-6}ns$) motivates the assumption that the measurement is performed at an \emph{arbitrarily} large time $T$. We prove that the dipole moment matrix is locally observable, that is, that for any two close but different dipoles, we can find a control which gives two distinct measurements to first order.

Identifying the dipole moment of a quantum system is important for fundamental problems in atomic physics such as the analysis of parity violation measurements \cite{Noecker-al-PRL_1988}. It can also serve as a test bed for theoretical calculations using novel techniques such as relativistic many body perturbation theory \cite{Safronova-al-PRA_1999}. Various methods have been considered for measuring dipole moment matrix elements (see \cite{Bayram-al-JPhysB_2006} and references therein). The most common method is to measure the lifetime of a quantum state \cite{Gomez-al-Orozco-PRA_2005}, which is related to the norm of the dipole moment matrix element by Fermi's golden rule. Although this procedure has been extensively used to extract matrix elements between the ground and first excited states, it can be difficult to generalize the method to measuring matrix elements between two excited states. Problems such as branching and super-radiance \cite{Dicke-PR_1954} complicate the interpretation of the data. Also, lifetime measurements only provide the norm of the matrix elements and give no information about the phases. Another commonly used method is multi-dimensional spectroscopy \cite{Tian-Warren-science_2003}, which can reveal valuable information on the dipole coupling structure of the system (i.e, revealing which dipole elements are zero and non-zero). However, multi-dimensional spectroscopy is experimentally challenging to realize and often does not give quantitative information about the dipole moments. Other approaches have also been investigated including the measurement of the degree of induced polarization as a function of the detuning in a two photon process \cite{Bayram-al-JPhysB_2006}, stark shift measurements \cite{Arora-al-PRA_2007} and Rayleigh, Raman scattering \cite{Havey-PhysLettA_1997}.

In this paper, we consider the identification of the dipole moment matrix as an optimal control problem \cite{Lu-Rabitz-JPhysChem_1995}. We prove the existence of controls which make the identification from one population measurement a well posed problem (theorem \ref{thm:observability}). The controls we construct for this purpose are inspired from the two pulses separated by a large time interval used in Ramsey interferometry (see, e.g.,~\cite[page 148]{Haroche-Raimond_2006}). In Figure~\ref{fig.pulse} we illustrate this analogy and explain how to achieve, in three steps,  a  control to discriminate  the real dipole moment between states $\ket{l}$ and $\ket{k}$. The field structure has the following components:
 \begin{enumerate}
 \item Steer the system from its known initial state to the state $\ket{l}$ with a pulse $\epsilon_1$.
 \item Perform a large number of Rabi flops between states $\ket{l}$ and $\ket{k}$ with a resonant control $\varepsilon\cos(\omega_{lk}(\tau-\tau_1))$, $\tau_1\ge\tau\le\tau_2$.
 \item Achieve a control pulse $\epsilon_2$ inducing an overlap with the measured population; such overlap will depend on the Rabi angle produced during the second step; thus the measured population will be sensitive to the dipole value between $\ket{l}$ and $\ket{k}$.
 \end{enumerate}
Such discriminating controls are exploited in proving the key lemma~\ref{lemma:gradientcontrollability} later in the paper.

The perspective above combined with control theory is motivated by three practical arguments. First, measuring a state population at one time $T$ is a technique which can have a very high signal to noise ratio ($\sim 100$). Second, technological progress with spatial light modulators (SLM) permits generating a broad variety of controls in the laboratory. Third, spectral phase interferometry for direct electric-field reconstruction of ultrashort optical pulses (SPIDER) \cite{Iaconis-Walmsley-OpticsLetters_1998} has proven to be a reliable technique to measure ultra-short pulsed fields. Hence, we are able to design a variety of controls and measure the actual controls created in the laboratory, along with the output signal. This scenario is suitable for parameter estimation, which is a well known problem in nonlinear system identification depending strongly on a careful  choice of discriminating control inputs.

Although the high potential of employing quantum control for Hamiltonian identification is clear, a full theoretical foundation is yet to be established. This effort is referred to as bilinear due to the form of the Schrodinger equation with the dipole and wave function entering multiplicatively. A vast literature exists for parameter estimation of bilinear systems, but to the best of our knowledge, identifiability results mainly rely on continuous time measurements linear with respect to the state. In a large set of quantum systems only the expectation value of a Hermitian operator can be measured, which is highly nonlinear with respect to the dipole. Le Bris and al \cite{LeBris-al-ESAIM_2007} prove the observability of the dipole moment matrix when the population of all states are measured over an arbitrarily large interval of time. Methods to reconstruct the dipole from the measured data were proposed using nonlinear observers \cite{Leghtas-Mirrahimi-Rouchon-CDC_2009,Bonnabel-al-Automatica_2009}. A different setting is considered in \cite{Schirmer-Langbein-ISCCSP_2010,Schirmer-Oi-LaserPhys_2010} where it is supposed that one can prepare and measure the system in a set of orthogonal states at various times, and the available data is the probability to measure the system in a certain state when it was prepared in another; Bayesian estimation is used to reconstruct the energy levels, the damping constants and the dipole moment from the measured data. We consider here the less demanding case where the only available measurement is the population of one state at one arbitrarily large time, and the initial state is  known and coincides with  the ground state.

The paper is organized as follows. In section \ref{sec:MainResult} we state the main result in Theorem \ref{thm:observability}, and section \ref{sec:proofs} gives the proof of the Theorem. The proof is based on a lemma which we prove in section \ref{sec:lemmaproof}.
Finally concluding remarks are presented in Section~\ref{sec:conclusion}.


\section{Observability of the quantum dipole moment}
\label{sec:MainResult}
\subsection{Problem setting}
\label{subsec:problemsetting}
We consider a quantum system in a pure state described by the wave function $\ket{\psi}\in\mathcal{S}$. Here $\mathcal{S}$ is the set of $N$ dimensional complex vectors of unit norm. The system interacts with an electric field (the real control input) $\epsilon\in\mathcal{E}_T$ for some $T>0$ with

\begin{equation}
\mathcal{E}_T = \mathcal{L}^1([0,T];\RR)= \left\{ f:[0,T] \to \RR \left|  \int_0^T |f(x)| dx < \infty \right. \right\}.
\end{equation}

\noindent For a given control $\epsilon$ we measure the population of the state $\ket{f}$ at time $T$ denoted as $P_{if}(\epsilon)$. We denote by $H_0$ the free Hamiltonian (Hermitian matrix) and
by $\mu$ the dipole moment operator, also a Hermitian matrix.  The initial state is an eigenvector of $H_0$ denoted $\ket{i}$. We consider a semi-classical model for the light-matter interaction \cite{Cohen-Tannoudji-al-AtomPhotonInteractions}, and the dynamics of $\ket{\psi}$ are given by the Schrodinger equation:
\begin{eqnarray}
\notag
\imath\hbar \dotex\ket{\psi(t)}&=&(H_0-\epsilon(t)\mu)\ket{\psi(t)}\\
\label{eq:dynamics_psi}
\ket{\psi(0)}&=&\ket{i}\\
\notag
\label{eq:measurement_psi}
P_{if}(\epsilon)&=&|\scprod{f}{\psi(T)}|^2
\end{eqnarray}
For all $T>0$, we suppose that we can create any field $\epsilon\in\mathcal{E}_T$ and that we can measure $P_{if}(\epsilon)$. For $M$ fields $\{\epsilon_1,..,\epsilon_M\}$ we can collect the measurements $\{P_{if}(\epsilon_1),..,P_{if}(\epsilon_M)\}$.
Through \eqref{eq:dynamics_psi}  $P_{if}$ is a function of $\mu$ and a functional of $\epsilon$, and when necessary this explicit dependence will be written as $P_{if}(\epsilon,\mu)$.
The aim of this paper is to explore the feasibility of estimating the dipole moment matrix $\mu$ from the measured data $\{P_{if}(\epsilon_1),..,P_{if}(\epsilon_M)\}$ using well chosen controls $\{\epsilon_1,..,\epsilon_M\}$. Below, $P_{if}(\epsilon,\mu)$ refers to the measurement achieved on the real system using a control $\epsilon$, and for any $\hat\mu\ne\mu$, $P_{if}(\epsilon,\hat\mu)$ is the estimated measurement which is obtained by simulating system \eqref{eq:dynamics_psi} with the control $\epsilon$
and coupling $\hat\mu$.

\subsection{Main result}
For all $k\le N$ we denote $\ket{k}$ as the eigenvector of $H_0$ with associated eigenvalue $E_k$. Throughout the paper, all matrices are written in the basis $\left(\ket{1},..,\ket{N}\right)$.
For all $k,l\le N$ we specify $\sigma^x_{lk}
\triangleq
\ket{l}\bra{k}+\ket{k}\bra{l}$. We define
\begin{equation}
\mathcal{M}\triangleq \text{Span}\{\sigma_x^{lk} \backslash k,l\le n \textrm{ such that } \tr{\mu\sigma_x^{lk}}\ne 0 \}, \ \ M = dim(\mathcal{M})
\end{equation}
with Tr being the trace operation. When all non diagonal elements of $\mu$ are non-null, $M=\frac{N(N-1)}{2}$.
The main result is the following:

\begin{thm}
\label{thm:observability}
Consider a real symmetric matrix $\mu$ with zero diagonal entries and a real diagonal matrix $H_0$ with non-degenerate transitions.
Suppose that the system state in \eqref{eq:dynamics_psi} is controllable. Then for any positive constant $\alpha$ there exists a time $T>0$ and $M$ fields
$(\epsilon_1,..,\epsilon_M)$ $\in {\mathcal{E}_{T}}^M$ such that the cost function
\begin{equation}
\label{eq:costfunction}
J: \hat\mu\in\mathcal{M}\rightarrow\sum_{k=1}^M{(P_{if}(\epsilon_k,\hat\mu)-P_{if}(\epsilon_k,\mu))^2}
\end{equation}
is locally $\alpha$-convex around $\mu$.
\end{thm}
In appendix \ref{appendix:def} we provide the definitions of controllability, $\alpha-$convexity and a matrix with non-degenerate transitions.\\
A direct consequence of Theorem \ref{thm:observability} is the local observability of the dipole moment matrix:
\begin{crl}
Under the assumptions of Theorem \ref{thm:observability}, the dipole moment matrix is locally observable.
\end{crl}
Here and throughout the paper, the norm of matrix $\mu$, noted $\norm{\mu}$ refers to the max norm.
\begin{proof}
Take $\alpha>0$. Theorem \ref{thm:observability} implies that there exists a time $T>0$ and $M$ fields $(\epsilon_1,..,\epsilon_M)$ $\in {\mathcal{E}_{T}}^M$ such that the cost function $J$ is locally $\alpha$-convex around $\mu$. Hence there $\exists r>0$ such that for all $\hat\mu\in \mathcal{M}$ with
$\norm{\hat\mu-\mu}\le r$ and $\hat\mu\ne\mu$, $J(\hat\mu)>0$, and hence there exists $\epsilon\in\{\epsilon_1,..,\epsilon_M\}$ such that $P_{if}(\epsilon,\hat\mu)-P_{if}(\epsilon,\mu)\ne 0$.\\
\end{proof}

\begin{rmk}
The local $\alpha-$convexity is a property stronger than the mere possibility to identify the dipole matrix. It states that the distinction between a dipole
candidate $\hat\mu$ and the true dipole $\mu$ can be observed (through the measurements aggregated in $J$) to first order in the distance $\| \mu - \hat\mu\|$. This first order dependence of the measurement $P_{if}$ with respect to the dipole $\mu$ is addressed in more detail in lemma \ref{lemma:gradientcontrollability}.
\end{rmk}

\begin{rmk}
The cost function defined in Theorem \ref{thm:observability} is locally $\alpha$ convex on $\mathcal{M}$. This suggests that all non-null matrix elements of $\mu$ can be reliably estimated with well chosen controls, but we cannot say the same for null matrix elements. In practice, when dealing with an atomic or molecular quantum system, it is generally well known which matrix elements are null by applying selection rules (see $A_{XIII}$, $1.c.\gamma$ in \cite{Cohen-al-MecaQ_1973}). Also, other methods, such as multi-dimensional spectroscopy, may be used to reveal the non-zero structure of the dipole moment matrix.
\end{rmk}

\subsection{Robustness}
This paper will not analyze the influence of noise effects on the estimation of the dipole moment. Some issues relevant to noise have been treated elsewhere \cite{Shir-al-IEEE_2010}. We recall the three main sources of uncertainties:
\begin{enumerate}
\item Uncertainty in the control $\epsilon(t)$: a predetermined control cannot be perfectly created in the laboratory, and the actually created control cannot be measured with arbitrary precision.
\item Uncertainty in the measured signal $P_{if}(\epsilon)$.
\item Uncertainty in the model: the bilinear Schrodinger equation \eqref{eq:dynamics_psi} is a simplified model not taking into account dissipation, collisions and higher order matter-field interactions.
\end{enumerate}
We consider a simple noise model in order to give an estimate of the robustness of employing quantum control for Hamiltonian identification. Suppose that instead of measuring $P_{if}(\epsilon_k,\mu)$, one actually measures $P_{if}(\epsilon_k,\mu)+\nu_k$, where $\nu_k$ is the realization of a stochastic variable $\nu$ of (finite) variance $var(\nu)$. We can then build a cost function $J_{noisy}$ with these measurement outcomes.
Then it can be proven that
$J_{noisy}$ will reach its minimum at a point $\mu_{noisy}$, which belongs to a ball centered around $\mu$ and of radius $R_{noise} \approx \sqrt{\frac{var(\nu)}{\alpha}}$ (with $\alpha$ defined in Theorem \ref{thm:observability}). This suggests that if the variance of the noise is a bounded function of the measurement time $T$ and the controls in $\mathcal{E}_T$, then the estimation of the dipole moment matrix can be arbitrarily precise by picking controls such that $\alpha$ is arbitrarily large.


\section{Proof of the Theorem \ref{thm:observability}}
\label{sec:proofs}

\subsection{Existence of discriminating controls}

Theorem \ref{thm:observability} is a local result and in order to prove it we need to study the first order variation of the measurement $P_{if}$ with respect to the dipole moment $\mu$ for a given time $T$ and control $\epsilon$. We denote $\mu'=\frac{1}{\norm{\mu}}\mu$ the normalized dimensionless dipole moment operator. We denote $\mu'_{lk}\triangleq\tr{\mu'\sigma_x^{lk}}$ and $\dv{P_{if}}{\mu'_{lk}}(\epsilon)$ the partial derivative of $P_{if}(\epsilon)$ with respect to $\mu'_{lk}$.

To prove Theorem \ref{thm:observability}, consider a class of controls which can be configured in terms of the physical framework sketched in Figure~\ref{fig.pulse} with a suggestive  analogy from Ramsey interferometry.
In experiments where shaped pulses are used as controls generated with spatial light modulators \cite{Rabitz-science_2003}, we can further expect that a \emph{good} pulse to identify a certain dipole moment $\mu_{lk}$ between states $\ket{l}$ and $\ket{k}$ can be achieved using the three steps set out in Section \ref{sec:introduction}.
In particular, theorem \ref{thm:observability} is based on the following lemma:
\begin{lm}
\label{lemma:gradientcontrollability}
Suppose that $\mu$ is real, symmetric and has only zeros on its diagonal and $H_0$ is real, diagonal, with non-degenerate transitions.
Suppose system \eqref{eq:dynamics_psi} is controllable. Then for all $(l,k)$ such that $\mu_{lk}\ne0$ there exists $\xi_0>0$ such that for all $\xi \in ]0,\xi_0[$ there exists $T>0$ and  $\epsilon\in\mathcal{E}_T$ such that
 \begin{enumerate}
 \item $\dv{P_{if}}{\mu'_{lk}}(\epsilon)=\frac{1}{2\xi}+O(1)$
 \item $\forall \{m,n\}\ne\{l,k\}$ $\dv{P_{if}}{\mu'_{mn}}(\epsilon)=O(1)$.
 \end{enumerate}
\end{lm}
Where $O(1)$ is a term bounded with respect to $\xi \in ]0,\xi_0[$.

\begin{rmk}
Lemma \ref{lemma:gradientcontrollability} claims that we can always find a time $T$ and a control $\epsilon$ which makes the data $P_{if}(\epsilon)$ arbitrarily sensitive to one element of $\mu$ while its sensitivity to the other matrix elements stays bounded. This is an existence result.
In the laboratory however we are constrained by the measurement time $T$ which has to be much smaller than the excited states lifetimes and the collision time as well as by the pulse energy and bandwidth which can not be arbitrarily large. These circumstances may result in the actual sensitivity being less than the theoretical estimate.
Besides, the model \eqref{eq:dynamics_psi} itself would possibly have to be modified for very strong fields \cite{Coron-al-NJP_2009} introducing extra unknown parameters in the dynamics.
\end{rmk}

\subsection{Proof of Theorem \ref{thm:observability}}
\begin{proof}
To each pair of integers $(l_p,k_p)$, $l_p<k_p$ such that $\tr{\mu'\sigma_x^{l_pk_p}}\ne 0$ we associate a unique index
$p\in \{1,...,M\}$, and we define $\sigma_x^p \triangleq \sigma_x^{l_pk_p}$ along with $\mu'_p \triangleq \tr{\mu'\sigma_x^p}$.

According to lemma $\ref{lemma:gradientcontrollability}$, $\exists \xi_0>0$ such that $\forall 0<\xi<\xi_0$ $\exists$ $T_1,..,T_M$ and $(\epsilon_1,..,\epsilon_M)\in{\mathcal{E}_{T_1}}\times..\times{\mathcal{E}_{T_M}}$ such that:

\begin{enumerate}
\item for all $p\in [1:M]$ $\dv{P_{if}}{\mu'_{p}}(\epsilon_p)=\frac{1}{2\xi}+O(1)$
\item $\forall p'\ne p$ $\dv{P_{if}}{\mu'_{p'}}(\epsilon_p)=O(1)$
\end{enumerate}

We take $T=\max(T_1,..,T_M)$ and for all $k\in\{1,..,M\}$ we extend the definition of $\epsilon_k$ from $[0,T_k]$ to $[0,T]$ by taking $\epsilon_k(t)=0$ for all $t\in]T_k,T]$.

We will use $J: \mathcal{M} \to \RR$ defined by:

\begin{equation*}
J(\hat\mu) = \sum_{k=1}^M{(P_{if}(\epsilon_k,\hat\mu)-P_{if}(\epsilon_k,\mu))^2} ,
\end{equation*}

and then find

\begin{eqnarray*}
\frac{\partial^2 J}{\partial\mu'_p\partial\mu'_{p'}}(\mu)&=&\sum_{k=1}^M{\frac{\partial P_{if}}{\partial \mu'_p}(\epsilon_k,\mu)\frac{\partial P_{if}}{\partial \mu'_{p'}}(\epsilon_k,\mu)}
\end{eqnarray*}

so that for all $p\in \{ 1, ..., M \} :  \frac{\partial^2 J}{\partial{\mu'_p}^2}(\mu)=\frac{1}{4\xi^2}+O(\frac{1}{\xi})$
and when $p\ne p'$ $\frac{\partial^2 J}{\partial\mu'_p\partial\mu'_{p'}}(\mu)=O(\frac{1}{\xi})$. We have:

\begin{equation*}
\nabla^2J(\mu)=\frac{1}{4\xi^2}\left(I +O(\xi)\right)
\end{equation*}
where $\nabla^2J(\mu)$ is the Hessian of $J$ at $\mu$ and $I$ is the identity matrix.
The smallest eigenvalue of $\nabla^2J(\mu)$ scales as
$\frac{1}{4\xi^2}(1+O(\xi))$, hence by taking $\xi$ small enough it can be made larger than $\alpha$ thereby reaching the conclusion above.

\end{proof}


\section{Proof of lemma \ref{lemma:gradientcontrollability}}
\label{sec:lemmaproof}
\subsection{Sketch of the proof:}
We introduce the dimensionless time scale $\tau \triangleq \frac{1}{\hbar}\norm{H_0}t$ and also $\top=\frac{1}{\hbar}\norm{H_0}T$.

\begin{center}
\begin{figure}[t]
\includegraphics[width=7.0in]{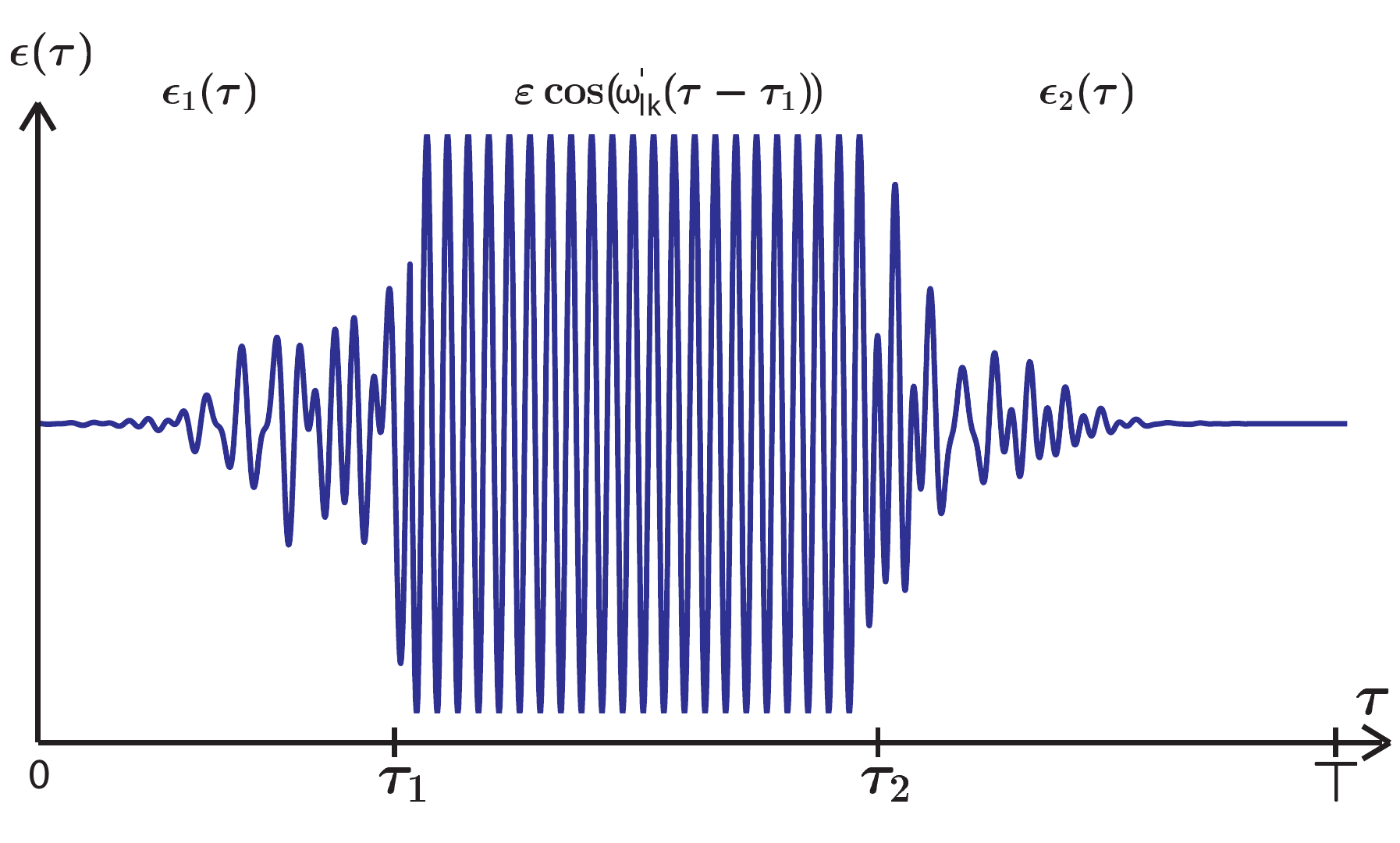}
\caption{A \emph{good} control $\epsilon$ has three components (inspired from Ramsey interferometry) to enable the identification of $\mu_{lk}$. The field $\epsilon_1$ is defined over $[0,\tau_1]$ (i.e, an analog of the first Ramsey pulse) to steer the known initial state $\ket i$ to $\ket{\psi_1}=\ket{l}$:  $\ket{l}=U(\tau_1,0)\ket{i}$. The field $\epsilon_2$ is defined over $[\tau_2,\top]$ (i.e, an analog of the second Ramsey pulse) is such that $\ket{f}=U(\top,\tau_2)\ket{\psi_2}$ where $\ket{\psi_2}$ is collinear to $\frac{\ket{l} + e^{\imath \phi}\ket{k}}{\sqrt{2}}$. The relative phase $\phi$ is such that $\ket{\psi_2}=U(\tau_2,\tau_1)\left(\frac{\ket{l}+ \imath \ket{k}}{\sqrt{2}}\right)$ where the propagator $U(\tau_2,\tau_1)$  corresponds, for a long interval $\tau_2-\tau_1$, to a large number of Rabi oscillations with  the control $\varepsilon \cos(\omega_{lk}^\prime(\tau-\tau_1))$ resonant with the $\ket{l}\leftrightarrow \ket{k}$ transition. }
\label{fig.pulse}
\end{figure}
\end{center}

In order to prove lemma \ref{lemma:gradientcontrollability}, we need to find a control $\epsilon\in\mathcal{E}$ which makes $\dv{P_{if}}{\mu'_{lk}}(\epsilon)$ arbitrarily large while $\dv{P_{if}}{\mu'_{mn}}(\epsilon)$ stays bounded. The concept is based on splitting the time interval $[0,\top]$ into three pieces $[0,\tau_1]$ $[\tau_1,\tau_2]$ and $[\tau_2,\top]$. Over $[\tau_1,\tau_2]$, playing the role of the time interval between the two Ramsey pulses, we will use a field resonant with the $l-k$ transition producing extensive Rabi flopping between states $\ket{l}$ and $\ket{k}$. However, in order for this Rabi flopping to contribute to increasing $\dv{P_{if}}{\mu'_{lk}}(\epsilon)$, we need to drive the known initial state $\ket{i}$ to a particular state $\ket{\psi_1}$ at $\tau=\tau_1$ (i.e, an analog of the first Ramsey pulse)  and also drive the final state $\ket{f}$ to a particular state $\ket{\psi_2}$ at $\tau=\tau_2$ (i.e, an analog of the second Ramsey pulse). By increasing $\tau_2-\tau_1$ (i.e an analog of the time interval between the two Ramsey pulses), we will see that $\dv{P_{if}}{\mu'_{lk}}(\epsilon)$ can be made arbitrarily large. Besides, $H_0$ having non-degenerate transitions, all other transitions will be out of resonance and hence we will show that as $\tau_2-\tau_1$ gets larger, $\forall \{m,n\}\ne\{l,k\}$ $\dv{P_{if}}{\mu'_{mn}}(\epsilon)$ stay bounded. The particular controls $\epsilon_1$ and $\epsilon_2$ over the time intervals $[0,\tau_1]$ and $[\tau_2,\top]$ are used to drive the system to the respective desired states $\ket{\psi_1}$ and $\ket{\psi_2}$ (see Figure \ref{fig.pulse}). Since we suppose the system to be controllable over the compact $\mathcal{S}$, we can find such controls which are defined over the \emph{bounded} time intervals $[0,\tau_1]$ and $[\tau_2,\top]$. Notice that such discriminating fields $\epsilon$ (plotted in Figure \ref{fig.pulse}) may be difficult to create in the laboratory due to their very specific spectrum (a peak at $\omega_{lk}$ and a broad distribution at other frequencies). This form of control is used here to construct an observability proof only. We are currently investigating the means to generate discriminating fields which may be readily created in the laboratory.

Before starting the proof we introduce some notation. For two times $\tau,\tau'\in [0,\top]$ we define the propagator $U(\tau',\tau)$ such that $\ket{\psi(\tau')}=U(\tau',\tau)\ket{\psi(\tau)}$. Rewriting \eqref{eq:dynamics_psi} for $U(\tau,0)$ we obtain:
\begin{eqnarray}
\label{eq:dynamics_U}
\imath \frac{\partial}{\partial\tau} U(\tau,0)&=&\frac{1}{\norm{H_0}}(H_0-\epsilon(\tau)\mu)U(\tau,0)\notag\\
U(0,0)&=&I\\
P_{if}(\epsilon)&=&|\bra{f}U(\top,0)\ket{i}|^2\notag
\end{eqnarray}

\subsection{Proof of lemma \ref{lemma:gradientcontrollability}}
\begin{proof}
The proof of lemma \ref{lemma:gradientcontrollability} has two parts I and II separately treated below.

\paragraph*{i) Part I}
Take two times $\tau_1,\tau_2$, $0<\tau_1<\tau_2<\top$. We can write (for any complex $z$ we denote by $\bar{z}$ its complex conjugate):
\begin{equation}
P_{if}(\epsilon)=|\bra{f}U(\top,\tau_2)U(\tau_2,\tau_1)U(\tau_1,0)\ket{i}|^2=z\bar z \textrm{ where }
z=\bra{f}U(\top,\tau_2)U(\tau_2,\tau_1)U(\tau_1,0)\ket{i}.
\end{equation}

The differential of $P_{if}$ is $\delta P_{if}=2\Re\left(\delta z \bar z\right)$ where $\Re()$ denotes the real part.

\begin{eqnarray}
\delta z &=& \bra{f}\delta U(\top,\tau_2)U(\tau_2,\tau_1)U(\tau_1,0)\ket{i}\nonumber\\
 &+&  \bra{f}U(\top,\tau_2)\delta U(\tau_2,\tau_1)U(\tau_1,0)\ket{i} \label{eq:dz}\\
 &+&  \bra{f}U(\top,\tau_2)U(\tau_2,\tau_1)\delta U(\tau_1,0)\ket{i}\nonumber
\end{eqnarray}

We find (see appendix \ref{appendix:calculation1})

\begin{equation}
\label{eq:dUdmulk}
\frac{\partial}{\partial \mu'_{lk}} U(\tau_2,\tau_1)=\imath\frac{\norm{\mu}}{\norm{H_0}} U(\tau_2,\tau_1)\int_{\tau_1}^{\tau_2}{\epsilon(\tau)U^\dag(\tau,\tau_1)\sigma_x^{lk}U(\tau,\tau_1)d\tau}
\end{equation}

Denote for any $m,n=1,...,M$ : $\omega'_{mn} \triangleq \frac{E_m-E_n}{\norm{H_0}}$ and consider the control defined on $[\tau_1,\tau_2]$:
 \begin{equation}
 \label{eq:control}
 \epsilon(\tau)=\varepsilon\cos(\omega'_{lk}(\tau-\tau_1))
 \end{equation}
where $\varepsilon$ is a small strictly positive real parameter. Take $\xi=\frac{\norm{\mu}\varepsilon}{\norm{H_0}}$. The only remaining degree of freedom in the
control over $[\tau_1,\tau_2]$ is $\xi$, which can be taken arbitrarily small. We define $H'_0=\frac{1}{\norm{H_0}}H_0$ and $\omega'_{mn}=\bra{m}H'_0\ket{m}-\bra{n}H'_0\ket{n}$. Note that $\omega'_{mn}=-\omega'_{nm}$.
We now rewrite \eqref{eq:dynamics_U} and \eqref{eq:dUdmulk} for the control given in \eqref{eq:control} on the time interval $[\tau_1,\tau_2]$:

\begin{eqnarray}
\label{eq:dynamics_U_new}
\imath \frac{\partial}{\partial\tau}U(\tau,\tau_1)&=&(H'_0-\xi\cos(\omega'_{lk}(\tau-\tau_1))\mu')U(\tau,\tau_1)
\end{eqnarray}

\begin{equation}
\label{eq:dUdmulk_new}
\frac{\partial}{\partial \mu'_{lk}} U(\tau_2,\tau_1)=\imath \xi U(\tau_2,\tau_1)\int_{\tau_1}^{\tau_2}{\cos(\omega'_{lk}(\tau-\tau_1))U^\dag(\tau,\tau_1)\sigma_x^{lk}U(\tau,\tau_1)d\tau}
\end{equation}

The goal is to show that $\frac{\partial}{\partial \mu'_{lk}} U(\tau_2,\tau_1)$ can be made arbitrarily "large" while $\frac{\partial}{\partial \mu'_{mn}} U(\tau_2,\tau_1)$ stays bounded. Note that all the terms in the integrand of \eqref{eq:dUdmulk_new} are bounded, and a rough estimate of the norm of $\frac{\partial}{\partial \mu'_{lk}} U(\tau_2,\tau_1)$ gives a quantity proportional to $(\tau_2-\tau_1)\xi$. Hence, we take $\tau_2-\tau_1= \frac{1}{\xi^2}$, implying the need to have expressions for $U(\tau,\tau_1)$ over a time scale on the order of $\frac{1}{\xi^2}$. To this end we state lemma \ref{lemma:approx} which gives such an approximation.

\begin{lm}
\label{lemma:approx}
Consider equation \eqref{eq:dynamics_U_new}. There exists a Hermitian matrix $K$ and $\xi_0>0$ such that for any $\xi \in ]0,\xi_0[$ and any $\tau\in[\tau_1,\tau_1+\frac{1}{\xi^2}]$ we have:
\begin{equation*}
U(\tau,\tau_1)=e^{-\imath H'_0(\tau-\tau_1)}e^{\imath(\xi \frac{\mu'_{lk}}{2}\sigma_x^{lk}+\xi^2 K)(\tau-\tau_1)}+O(\xi)\; ,
\end{equation*}
where $O(\xi)$ is such that $\frac{1}{\xi}O(\xi)$ is bounded and does not depend on $\tau$.
\end{lm}

We continue with the proof of Lemma~\ref{lemma:gradientcontrollability} and will come back to Lemma~\ref{lemma:approx} in
Section~\ref{subsec:proof2}.

Using the expression of $U(\tau,\tau_1)$ given in lemma \ref{lemma:approx}, the integrand in \eqref{eq:dUdmulk_new} is:
\begin{eqnarray*}
&&\cos(\omega'_{lk}(\tau-\tau_1))U^\dag(\tau,\tau_1)\sigma_x^{lk}U(\tau,\tau_1)=\\
&&e^{-\imath(\xi \frac{\mu'_{lk}}{2} \sigma_x^{lk}+\xi^2 K)(\tau-\tau_1)}\left(\cos(\omega'_{lk}(\tau-\tau_1))e^{\imath H'_0(\tau-\tau_1)}\sigma_x^{lk}e^{-\imath H'_0(\tau-\tau_1)}\right)e^{\imath(\xi \frac{\mu'_{lk}}{2} \sigma_x^{lk}+\xi^2 K)(\tau-\tau_1)} + O(\xi)
\end{eqnarray*}
In order to compute \eqref{eq:dUdmulk_new}, we need the following result:
\begin{eqnarray}
\label{eq:transsigmaxlk}
\cos(\omega'_{lk}(\tau-\tau_1))e^{\imath H'_0(\tau-\tau_1)}\sigma_x^{lk}e^{-\imath H'_0(\tau-\tau_1)}&=&\frac{1}{2}\sigma_x^{lk}\\
&+&\frac{1}{2}\cos(2\omega'_{lk}(\tau-\tau_1))\sigma_x^{lk}+\frac{1}{2}\sin(2\omega'_{lk}(\tau-\tau_1))\sigma_y^{lk}\; ,\notag
\end{eqnarray}
where we denote $\sigma_y^{lk}=-\imath \ket{l}\bra{k}+\imath \ket{k}\bra{l}$.
In \eqref{eq:transsigmaxlk}, the terms oscillating at frequency $2\omega'_{lk}$ independent of $\xi$
will only contribute to $O(\xi)$ in \eqref{eq:dUdmulk_new}. We now focus on the contribution of the term with $\sigma_x^{lk}$ in \eqref{eq:dUdmulk_new} calling for $e^{-\imath(\xi \frac{\mu'_{lk}}{2}\sigma_x^{lk}+\xi^2 K)(\tau-\tau_1)}\sigma_x^{lk}e^{\imath(\xi \frac{\mu'_{lk}}{2}\sigma_x^{lk}+\xi^2 K)(\tau-\tau_1)}$. We find: (see appendix \ref{appendix:calculation2}) $\forall \tau$
  \begin{equation}
  \label{eq:transsigmaxlk2}
  e^{-\imath(\xi \frac{\mu'_{lk}}{2}\sigma_x^{lk}+\xi^2 K)(\tau-\tau_1)}\sigma_x^{lk}e^{\imath(\xi \frac{\mu'_{lk}}{2}\sigma_x^{lk}+\xi^2 K)(\tau-\tau_1)}=\sigma_x^{lk}+O(\xi)
  \end{equation}

Introducing \eqref{eq:transsigmaxlk2} into \eqref{eq:dUdmulk_new}, we find:

\begin{equation}
\frac{\partial}{\partial \mu'_{lk}} U(\tau_2,\tau_1)=\imath \xi U(\tau_2,\tau_1)\left(\frac{\tau_2-\tau_1}{2}\sigma_x^{lk}+O(1)+(\tau_2-\tau_1)O(\xi)\right).
\end{equation}

From now on, we take $\tau_2=\tau_1+\frac{1}{\xi^2}$ and obtain:

\begin{equation}
\label{eq:dUdmulkapprox}
\frac{\partial}{\partial \mu'_{lk}} U(\tau_2,\tau_1)=\imath U(\tau_2,\tau_1)(\frac{1}{2\xi}\sigma_x^{lk}+O(1)).
\end{equation}

\noindent We define $\ket{\psi_1}\triangleq\ket{l}$ and
$\ket{\psi_2}\triangleq\frac{1}{\sqrt{2}}U(\tau_2,\tau_1)(\ket{l}+ \imath \ket{k})$. Since the system is controllable there exists a time $\tau_1$ and a field $\epsilon_1\in\mathcal{E}_{\tau_1}$ such that $U(\tau_1,0)\ket{i}=\ket{\psi_1}$, and there exists a time $\top$ and  a field $\epsilon_2$ defined over $[\tau_2,\top]$ such that $U^\dag(\top,\tau_2)\ket{f}=\ket{\psi_2}$.\\

Since the state space is compact (here it is a sphere), we know that if the system is controllable, it is controllable in bounded time, and with bounded controls (see Theorem 6.5 in \cite{Jurdjevic-Sussmann-JDiffEq_1972}). We can therefore define a bounded set of times $\{T(\psi_i,\psi_f)\backslash (\psi_i,\psi_f)\in\mathcal{S}^2\}$ and a bounded set of controls (bounded for the norm $\mathcal{L}^\infty$) $\{\epsilon(\psi_i,\psi_f)\in\mathcal{E}_{T(\psi_i,\psi_f)}\backslash (\psi_i,\psi_f)\in\mathcal{S}^2\}$ such that for any two states $\psi_i$ and $\psi_f$ in $\mathcal{S}$ the solution of~\eqref{eq:dynamics_psi} with the control $\epsilon(\psi_i,\psi_f)$
and starting from the initial state $\psi(0)=\psi_i$ arrives in $\psi_f$ at time $T(\psi_i,\psi_f)$.
Hence, $\top-\tau_2$ can be chosen bounded for all $\xi$.
Therefore $\frac{\partial}{\partial \mu_{lk}}U(0,\tau_1)$ and $\frac{\partial}{\partial \mu_{lk}}U(\tau_2,\top)$ are bounded. Thus, we have:
\small
\begin{equation}
\label{eq:dPifdmulk}
\frac{\partial}{\partial \mu'_{lk}}P_{if}(\epsilon)=2\Re(\bra{f}U(\top,\tau_2)\frac{\partial}{\partial \mu'_{lk}}U(\tau_2,\tau_1)U(\tau_1,0)\ket{i}\bra{i}U^\dag(\tau_1,0) U^\dag(\tau_2,\tau_1) U^\dag(\top,\tau_2)\ket{f})+O(1)
\end{equation}
\normalsize
We now utilize $U(\tau_1,0)\ket{i}=\ket{\psi_1}$ and $U^\dag(\top,\tau_2)\ket{f}=\ket{\psi_2}$ where $\ket{\psi_1}$ and $\ket{\psi_2}$ are defined above, and replace $\frac{\partial}{\partial \mu'_{lk}}U(\tau_1,\tau_2)$ by its expression in \eqref{eq:dUdmulkapprox} to find:

\begin{equation*}
\frac{\partial}{\partial \mu'_{lk}}P_{if}(\epsilon)=\frac{1}{2\xi}+O(1).
\end{equation*}

\noindent This expression holds for the control defined as:
\begin{equation}
\label{eq:controlopt}
\epsilon(\tau) =
\left\{
  \begin{array}{ll}
        \epsilon_1(\tau), & \text{ if } \tau\in[0,\tau_1] \\
        \frac{\norm{H_0}}{\norm{\mu}}\xi\cos(\omega'_{lk}(\tau-\tau_1)), & \text{ if } \tau\in]\tau_1,\tau_2[ \\
        \epsilon_2(\tau), & \text{ if } \tau\in[\tau_2,\top]
  \end{array}
\right.
\end{equation}

\paragraph*{ii) Part II}

We now need to prove that $\frac{\partial}{\partial \mu'_{mn}}P_{if}(\epsilon)=O(1)$ for $\{m,n\}\ne\{l,k\}$, where $\epsilon$ is the control found above in \eqref{eq:controlopt}.
As in equation \eqref{eq:dUdmulk_new}, we have:

\begin{equation}
\label{eq:dUdmumn}
\frac{\partial}{\partial \mu'_{mn}} U(\tau_2,\tau_1)=\imath \xi U(\tau_2,\tau_1)\int_{\tau_1}^{\tau_2}{\cos(\omega'_{lk}(\tau-\tau_1))U^\dag(\tau,\tau_1)\sigma_x^{mn}U(\tau,\tau_1)d\tau}\; ,
\end{equation}

and again the result of lemma \ref{lemma:approx} is employed. Equation \eqref{eq:dUdmumn} calls for

\begin{eqnarray}
\label{eq:transsigmaxmn}
&&\cos(\omega'_{lk}(\tau-\tau_1))e^{\imath H'_0(\tau-\tau_1)}\sigma_x^{mn}e^{-\imath H'_0(\tau-\tau_1)}=\notag\\
&&\frac{1}{2}\cos((\omega'_{lk}-\omega'_{mn})(\tau-\tau_1))\sigma_x^{mn}-\frac{1}{2}\sin((\omega'_{lk}-\omega'_{mn})(\tau-\tau_1))\sigma_y^{mn}\\
&&+\frac{1}{2}\cos((\omega'_{lk}+\omega'_{mn})(\tau-\tau_1))\sigma_x^{mn}+\frac{1}{2}\sin((\omega'_{lk}+\omega'_{mn})(\tau-\tau_1))\sigma_y^{mn}\; .\notag
\end{eqnarray}

Considering that $H_0$ has non-degenerate transitions (see definition in appendix \ref{appendix:def}) implies that $\omega'_{lk}-\omega'_{mn}\ne 0$ and $\omega'_{lk}+\omega'_{mn}\ne 0$. As the expression in \eqref{eq:transsigmaxmn} oscillates at frequencies independent of $\xi$, it therefore contributes to $O(\xi)$ in \eqref{eq:dUdmumn}. Hence, for $\tau_2-\tau_1=\frac{1}{\xi^2}$ we can directly conclude that:
\begin{equation*}
\frac{\partial}{\partial \mu'_{mn}} P_{if}(\epsilon)=O(1)\;.
\end{equation*}
\end{proof}

\subsection{Proof of lemma \ref{lemma:approx}} \label{subsec:proof2}
\begin{proof}
This proof relies on three consecutive changes of frame that aim to cancel the oscillating terms of order $1$ and $\xi$ in the dynamics. We then derive a specific form of the averaging Theorem (see theorem 4.3.6 in \cite{Sanders-Verhulst-Murdock-Spinger_2007} for a general form of the averaging theorem).\\

For the sake of clarity and with no loss of generality, we take $\tau_1=0$ and note $U(\tau)\triangleq U(\tau,\tau_1)$.
Equation \eqref{eq:dynamics_U_new} may be written in the interaction frame $U_I(\tau) \triangleq e^{\imath H'_0 \tau}U(\tau)$,
\begin{eqnarray*}
\label{eq:dynamics_UI}
\frac{\partial}{\partial\tau} U_I(\tau)&=&\imath\xi (\frac{\mu'_{lk}}{2}\sigma_x^{lk}+\frac{\partial}{\partial\tau}H_I(\tau))U_I(\tau)
\end{eqnarray*}
where:
\begin{eqnarray}
\label{eq:dHIdtau}
\frac{\partial}{\partial\tau}H_I(\tau)&=&\frac{1}{2}\sum_{(m,n)\ne(k,l)}\mu'_{mn}e^{\imath(-\omega'_{kl}+\omega'_{mn})\tau}\ket{m}\bra{n}
\\
&+& \frac{1}{2}\sum_{(m,n)\ne(l,k)}\mu'_{mn}e^{\imath(-\omega'_{lk}+\omega'_{mn})\tau}\ket{m}\bra{n}\notag
\end{eqnarray}
and the average of $H_I$ is zero. The average of a time dependent operator $O(\tau)$ is defined as follows (see defintion 4.2.4 in \cite{Sanders-Verhulst-Murdock-Spinger_2007}):
\begin{equation}
\label{eq:average}
\overline{O}=\lim_{\theta\rightarrow+\infty}\frac{1}{\theta}\int_0^\theta{O(\tau)d\tau}
\end{equation}

We now take $U'_{I}(\tau)=(I-\imath\xi H_I(\tau))U_I(\tau)$. Since $\frac{\partial}{\partial\tau}H_I$ is almost periodic (cf. appendix \ref{appendix:def} for the definition), then $H_I$ is also almost periodic and hence bounded for all $\tau$. Hence, there exists $\xi_0>0$, $\forall \xi<\xi_0$, and we have that $I-\imath\xi H_I(\tau)$ has an inverse and $(I-\imath\xi H_I(\tau))^{-1}=I+\imath\xi H_I(\tau)+O(\xi^2)$. We find:
\begin{equation*}
\frac{\partial}{\partial\tau}U'_I(\tau)=\imath\left(\xi \frac{\mu'_{lk}}{2}\sigma_x^{lk}-\imath\xi^2\left(\frac{\mu'_{lk}}{2}[H_I(\tau),\sigma_x^{lk}]+H_I(\tau)\frac{\partial}{\partial\tau}H_I(\tau)\right)+O(\xi^3)\right)U'_I(\tau)
\end{equation*}
$\frac{\mu'_{lk}}{2}[H_I(\tau),\sigma_x^{lk}]+H_I(\tau)\frac{\partial}{\partial\tau}H_I(\tau)=\imath(K+\frac{\partial}{\partial\tau}\tilde K(\tau))$ where $K=-\imath\overline{H_I\frac{\partial}{\partial\tau}H_I}$ does not depend on $\xi$, and $\tilde K(\tau)$ is almost periodic of average zero and bounded for all $\tau$. It is important to note that $\frac{1}{2}\overline{\frac{\partial}{\partial\tau}H_I^2}=0=\overline{H_I\frac{\partial}{\partial\tau}H_I}+\overline{(\frac{\partial}{\partial\tau}H_I)H_I}=\imath(K-K^\dag)$. Hence $K=K^\dag$ is Hermitian.

We now take $U''_I(\tau)=(I-\imath\xi^2\tilde K(\tau))U'_I(\tau)$. Since $\tilde K(\tau)$ is bounded for all $\tau$, then for a sufficiently small $\xi$, $I-\imath\xi^2\tilde K(\tau)$ has an inverse and $(I-\imath\xi^2\tilde K(\tau))^{-1}=I+\imath\xi^2\tilde K(\tau)+O(\xi^4)$. $U''_I$ satisfies the following equation:

\begin{equation}
\label{eq:dynamics_U''}
\frac{\partial}{\partial\tau}U''_I(\tau)=\imath\left(\xi \frac{\mu'_{lk}}{2}\sigma_x^{lk}+\xi^2K+O(\xi^3)\right)U''_I(\tau)\; ,
\end{equation}

and we define $U_{av}$ to be the solution to the averaged dynamics:
\begin{eqnarray}
\label{eq:dynamics_Uav}
\frac{\partial}{\partial\tau}U_{av}(\tau)&=&\imath\left(\xi \frac{\mu'_{lk}}{2}\sigma_x^{lk}+\xi^2K\right)U_{av}(\tau)\\
U_{av}(0)&=&I\notag \; .
\end{eqnarray}

We can directly solve \eqref{eq:dynamics_Uav}:
\begin{equation*}
U_{av}(\tau)=e^{\imath\left(\xi \frac{\mu'_{lk}}{2}\sigma_x^{lk}+\xi^2K\right)\tau}\; .
\end{equation*}
Subtracting \eqref{eq:dynamics_U''} from \eqref{eq:dynamics_Uav}, we find, using Gronwall's lemma, that for all $\tau<\frac{1}{\xi^2}$
one has $U''_I(\tau)=U_{av}(\tau)+O(\xi)$. Also note that to go from $U_I$ to $U''_I$ we have used two consecutive changes of variables which are close to the identity,
hence: $\forall \tau$ $U''_I(\tau)=U_I(\tau)+O(\xi)$. Finally, since $e^{-\imath H'_0\tau}$ is an isometry, we have:
\begin{equation*}
U(\tau)=e^{-\imath H'_0\tau}e^{\imath(\xi \frac{\mu'_{lk}}{2}\sigma_x^{lk}+\xi^2 K)\tau}+O(\xi)
\textrm{ for all } \tau\le\frac{1}{\xi^2}.
\end{equation*}
\end{proof}


\section{Conclusion}
\label{sec:conclusion}
We have proved that the identification of the real dipole moment matrix of a controllable
finite dimensional quantum system with non-degenerate transitions using as measurements only one population at a final time $T$ is a well posed problem. That is,
we can always find a control which discriminates between two close but different dipole moment matrices leading to two different measurements
and this distinction is of first order.
What at first may appear to be very restrictive data can actually be turned into a rich source of information by adequately controlling the dynamics of the system. \\
It would be interesting to extend this
result to the more general case of a Hermitian observable $O$ besides the population projection operator. Also, the extension to the case where the dipole moment $\mu$ is an arbitrary Hermitian operator (not necessarily real with a nul diagonal) would be another important step. Finally, further directions include establishing under what conditions the identification is globally well posed (i.e, not just locally) and exploring the impact of laboratory noise. We are now investigating the experimental implementation of a Hamiltonian identification procedure along the lines of the proposal given in \cite{Geremia-Rabitz-PRL_2002}, where the aim is to find, in the laboratory, those \emph{good} sets of controls $\{\epsilon_1,..,\epsilon_m\}$ complying with all experimental constraints which make the inversion procedure well posed.\\

\section*{Acknowledgment}
We thank Mazyar Mirrahimi for discussions. ZL acknowledges support from Agence Nationale
de la Recherche (ANR), Projet Jeunes Chercheurs EPOQ2 number ANR-09-JCJC-0070. GT was partially supported by the ANR, Project C-QUID BLAN-3-139579 and by the CNRS through a PICS grant. PR acknowledges support from ANR, Project C-QUID BLAN-3-139579. We acknowledge support from the US department of energy.

\bibliography{bibliography}


\appendices

\section{Definitions}
\label{appendix:def}
In this appendix, we give some useful definitions:

\begin{dfn}
\label{dfn:controllability}
We say that system \eqref{eq:dynamics_psi} is controllable\cite{Ramakrishna-al_PRA_1995} if and only if for all $\ket{\psi_1},\ket{\psi_2}\in\mathcal{S}$ there exists a time $t$ and a control $\epsilon\in\mathcal{E}_t$ such that for $\ket{\psi(0)}=\ket{\psi_1}$ \eqref{eq:dynamics_psi} leads to $\ket{\psi(t)}=\ket{\psi_2}$.
\end{dfn}

\begin{dfn}
We define a function $J:\hat\mu\in\mathcal{M}\rightarrow J(\hat\mu)\in\RR$. Let $\alpha>0$. Suppose $J$ is $\mathcal{C}^2$.
Let $\nabla^2 J(\hat\mu)$ be the Hessian of $J$ at $\hat\mu$. We say that $J$ is locally $\alpha$-convex \cite{Allaire_2007} around $\mu$ if the smallest eigenvalue of $\nabla^2 J(\mu)$
is larger than $\alpha$.
\end{dfn}

\begin{dfn}
Let $H_0$ be a real $N\times N$ diagonal matrix with entries $E_1,..,E_N$. We say that $H_0$ has non-degenerate transitions \cite{Turinici-Rabitz-JPhysA_2003} if $\forall (l,k)\ne (m,n)$ $l\ne k$ and $m\ne n$, we have $\left|E_l-E_k\right|\ne \left|E_m-E_n\right|$.
\end{dfn}

\begin{dfn}
Take system \eqref{eq:dynamics_psi}. Let us denote $\mathcal{M}$ as the space to which $\mu$ belongs. We say that $\mu$ is locally observable in $\mathcal{M}$
if there exists $r>0$ such that for all $\hat\mu\in\mathcal{M}$ with $0< \norm{\hat\mu-\mu}\le r$  there exists $T>0$ and $\epsilon\in\mathcal{E}_T$ such that $P_{if}(\epsilon,\hat\mu)\ne P_{if}(\epsilon,\mu)$.
\end{dfn}

For more details on almost periodic functions, one can refer to \cite{Bohr_1947}. Here, we give a very simplistic definition which is suitable for our needs.
\begin{dfn}
We say that an operator $A(t)$ is almost periodic if there exists $M\in\NN$ and $M$ time independent operators $A_0,..,A_M$ and $M$ frequencies $\omega_1,..,\omega_M$ such that for all $t$: $A(t)=\sum_{k=1}^M{e^{i\omega_k t}A_k}$.
\end{dfn}

\section{Calculations}
\subsection{Calculation 1}
\label{appendix:calculation1}
We desire to compute $\delta U(\tau_2,\tau_1)$ \cite{Beltrani-al-JChemPhys_2007} with only $\mu$ varying, and all other elements in the dynamics are fixed. Taking the differential of equation \eqref{eq:dynamics_U}, we get:

\begin{equation}
\label{eq:dynamics_dU}
\frac{\partial}{\partial\tau_2} \delta U(\tau_2,\tau_1)=-\frac{\imath}{\norm{H_0}}((H_0-\epsilon(\tau_2)\mu)\delta U(\tau_2,\tau_1)-\epsilon(\tau_2)\delta\mu U(\tau_2,\tau_1))\; .
\end{equation}
We seek a solution of \eqref{eq:dynamics_dU} in the form $\delta U(\tau_2,\tau_1)=U(\tau_2,\tau_1)\beta(\tau_2)$ where $\beta$ is a matrix to be found satisfying $\beta(\tau_1)=0$. Substituting this in \eqref{eq:dynamics_dU}, we find:

\begin{equation*}
\beta(\tau_2)=\imath\frac{1}{\norm{H_0}}\int_{\tau_1}^{\tau_2}{\epsilon(\tau)U^\dag(\tau,\tau_1)\delta\mu U(\tau,\tau_1)d\tau}
\end{equation*}

Hence, we have:
\begin{equation*}
\label{eq:deltaU}
\delta U(\tau_2,\tau_1)=\imath\frac{1}{\norm{H_0}} U(\tau_2,\tau_1)\int_{\tau_1}^{\tau_2}{\epsilon(\tau)U^\dag(\tau,\tau_1)\delta\mu U(\tau,\tau_1)d\tau}
\end{equation*}

and in particular:

\begin{equation}
\frac{\partial}{\partial \mu'_{lk}} U(\tau_2,\tau_1)=\imath\frac{\norm{\mu}}{\norm{H_0}} U(\tau_2,\tau_1)\int_{\tau_1}^{\tau_2}{\epsilon(\tau)U^\dag(\tau,\tau_1)\sigma_x^{lk}U(\tau,\tau_1)d\tau}
\end{equation}

\subsection{Calculation 2}
\label{appendix:calculation2}
In this section, we compute $e^{-i(\xi \frac{\mu'_{lk}}{2}\sigma_x^{lk}+\xi^2 K)(\tau-\tau_1)}\sigma_x^{lk}e^{i(\xi \frac{\mu'_{lk}}{2}\sigma_x^{lk}+\xi^2 K)(\tau-\tau_1)}$. Since we assumed $\mu'_{lk}\ne 0$,
$$
e^{-i(\xi \frac{\mu'_{lk}}{2}\sigma_x^{lk}+\xi^2 K)(\tau-\tau_1)}\sigma_x^{lk}e^{i(\xi \frac{\mu'_{lk}}{2}\sigma_x^{lk}+\xi^2 K)(\tau-\tau_1)}=e^{-i(\sigma_x^{lk}+\xi \frac{2K}{\mu'_{lk}})\frac{\mu'_{lk}}{2}\xi (\tau-\tau_1)}\sigma_x^{lk}e^{i( \sigma_x^{lk}+\xi \frac{2K}{\mu'_{lk}})\frac{\mu'_{lk}}{2}\xi (\tau-\tau_1)}
$$
$\sigma_x^{lk}+\xi \frac{2K}{\mu'_{lk}}$ is Hermitian, and is therefore diagonalizable. Hence, there exists a unitary matrix $P_\xi$ and a real diagonal matrix $\Delta_\xi$ such that $\sigma_x^{lk}+\xi \frac{2K}{\mu'_{lk}}=P_\xi\times \Delta_\xi\times P_\xi^\dag$. The function $\xi\in[0,\xi_0]\rightarrow \sigma_x^{lk}+\xi \frac{2K}{\mu'_{lk}}$ is analytic, therefore the eigenvectors of $\sigma_x^{lk}+\xi \frac{2K}{\mu'_{lk}}$ can be continued analytically as a function of $\xi$ (see Theorem 6.1 in chapter II, $\S6$ section 1 in \cite{Kato_1966}). Hence, we can write $P_\xi=P_0+O(\xi)$ where $P_0$ is such that $P_0\times\sigma_x^{lk}\times P_0^\dag=\sigma_z^{lk}$ is real and diagonal. $\sigma_z^{lk}=\proj{l}-\proj{k}$. We have, $\forall \tau$

\begin{eqnarray*}
e^{-i(\sigma_x^{lk}+\xi \frac{2K}{\mu'_{lk}})\frac{\mu'_{lk}}{2}\xi (\tau-\tau_1)}\sigma_x^{lk}e^{i( \sigma_x^{lk}+\xi \frac{2K}{\mu'_{lk}})\frac{\mu'_{lk}}{2}\xi (\tau-\tau_1)}&=&P_\xi e^{-i\frac{\mu'_{lk}}{2}\xi\Delta_\xi(\tau-\tau_1)} P_\xi^\dag\sigma_x^{lk}P_\xi e^{i\frac{\mu'_{lk}}{2}\xi\Delta_\xi(\tau-\tau_1)} P_\xi^\dag\\
&=&P_0 e^{-i\frac{\mu'_{lk}}{2}\xi\Delta_\xi(\tau-\tau_1)} P_0^\dag\sigma_x^{lk}P_0 e^{i\frac{\mu'_{lk}}{2}\xi\Delta_\xi(\tau-\tau_1)} P_0^\dag+O(\xi)\\
&=&P_0 e^{-i\frac{\mu'_{lk}}{2}\xi\Delta_\xi(\tau-\tau_1)} \sigma_z^{lk} e^{i\frac{\mu'_{lk}}{2}\xi\Delta_\xi(\tau-\tau_1)} P_0^\dag+O(\xi)\\
&=&P_0 \sigma_z^{lk} P_0^\dag+O(\xi)\\
&=&\sigma_x^{lk}+O(\xi)
\end{eqnarray*}

\end{document}